\documentclass[12pt]{article}

\usepackage{fancyhdr}
\usepackage{aaspp4}
\usepackage{epsf}
\usepackage{flushrt}
\usepackage{xfrac}
\usepackage{graphicx, afterpage}
\usepackage{listings}
\usepackage{color}
\usepackage{colortbl}
\usepackage{longtable,threeparttablex}
\usepackage[tableposition=top]{caption}
\usepackage{float}
\usepackage{indentfirst}
\usepackage{subfigure}
\usepackage[format=plain, labelsep=none, justification=justified, font=footnotesize, labelfont=it, singlelinecheck=false]{caption}

\usepackage{natbib}
\usepackage[colorlinks=true,citecolor=blue,urlcolor=cyan]{hyperref}
\newcommand\degree{\degr}
\newcommand\degrees\degree

\newcommand\hst{\em HST}

\DeclareSymbolFont{UPM}{U}{eur}{m}{n}
\DeclareMathSymbol{\umu}{0}{UPM}{"16}
\let\oldumu=\umu
\renewcommand\umu{\ifmmode\oldumu\else\math{\oldumu}\fi}
\newcommand\micro{\umu}
\renewcommand\micron{\micro m}
\newcommand\microns \micron

\let\oldsim=\sim
\renewcommand\sim{\ifmmode\oldsim\else\math{\oldsim}\fi}
\let\oldpm=\pm
\renewcommand\pm{\ifmmode\oldpm\else\math{\oldpm}\fi}
\newcommand\by{\ifmmode\times\else\math{\times}\fi}

\newbox{\wdbox}
\renewcommand\c{\setbox\wdbox=\hbox{,}\hspace{\wd\wdbox}}
\renewcommand\i{\setbox\wdbox=\hbox{i}\hspace{\wd\wdbox}}

\newcount\timect
\newcount\hourct
\newcount\minct
\newcommand\now{\timect=\time \divide\timect by 60
         \hourct=\timect \multiply\hourct by 60
         \minct=\time \advance\minct by -\hourct
         \number\timect:\ifnum \minct < 10 0\fi\number\minct}

\catcode`@=11

\newcommand\comment[1]{}
\newcommand\commenton{\catcode`\%=14}
\newcommand\commentoff{\catcode`\%=12}

\renewcommand\math[1]{$#1$}
\newcommand\mathshifton{\catcode`\$=3}
\newcommand\mathshiftoff{\catcode`\$=12}

\comment{the backslash is necessary}

\comment{alignment tab}

\let\atab=&
\newcommand\atabon{\catcode`\&=4}
\newcommand\ataboff{\catcode`\&=12}

\let\oldmsp=\sp
\let\oldmsb=\sb
\def\sp#1{\ifmmode
           \oldmsp{#1}%
         \else\strut\raise.85ex\hbox{\scriptsize #1}\fi}
\def\sb#1{\ifmmode
           \oldmsb{#1}%
         \else\strut\raise-.54ex\hbox{\scriptsize #1}\fi}
\newbox\@sp
\newbox\@sb
\def\sbp#1#2{\ifmmode%
           \oldmsb{#1}\oldmsp{#2}%
         \else
           \setbox\@sb=\hbox{\sb{#1}}%
           \setbox\@sp=\hbox{\sp{#2}}%
           \rlap{\copy\@sb}\copy\@sp
           \ifdim \wd\@sb >\wd\@sp
             \hskip -\wd\@sp \hskip \wd\@sb
           \fi
        \fi}
\def\msp#1{\ifmmode
           \oldmsp{#1}
         \else \math{\oldmsp{#1}}\fi}
\def\msb#1{\ifmmode
           \oldmsb{#1}
         \else \math{\oldmsb{#1}}\fi}
\def\supon{\catcode`\^=7}
\def\supoff{\catcode`\^=12}
\def\subon{\catcode`\_=8}
\def\suboff{\catcode`\_=12}
\def\supsubon{\supon \subon}
\def\supsuboff{\supoff \suboff}

\newcommand\actcharon{\catcode`\~=13}
\newcommand\actcharoff{\catcode`\~=12}

\newcommand\paramon{\catcode`\#=6}
\newcommand\paramoff{\catcode`\#=12}

\comment{And now to turn us totally on and off...}

\newcommand\reservedcharson{\commenton \mathshifton \atabon \supsubon \actcharon
	\paramon}

\newcommand\reservedcharsoff{\commentoff \mathshiftoff \ataboff
	\supsuboff \actcharoff \paramoff}

\catcode`@=12
\reservedcharsoff

\reservedcharson

\comment{ Must have ONLY ONE of these... trust these macros, they work

}

\newenvironment{packed_enum}{
\begin{enumerate}
   \setlength{\itemsep}{1pt}
   \setlength{\parskip}{0pt}
   \setlength{\parsep}{0pt}
}{\end{enumerate}}	

\newcommand{\squishlist}{
 \begin{list}{$\bullet$}
  { \setlength{\itemsep}{1pt}
     \setlength{\parsep}{0pt}
     \setlength{\topsep}{3pt}
     \setlength{\partopsep}{0pt}
     \setlength{\leftmargin}{2.0em}
     \setlength{\labelwidth}{1.5em}
     \setlength{\labelsep}{0.5em} } }

\newcommand{\squishend}{
  \end{list}  }

\reservedcharson
\actcharon

\reservedcharson
\actcharon

\usepackage{color}
\definecolor{dkgreen}{rgb}{0,0.6,0}
\definecolor{gray}{rgb}{0.5,0.5,.5}
\definecolor{mauve}{rgb}{0.58,0,0.82}
\definecolor{lgray}{rgb}{0.9,0.9,0.9}

\usepackage{listings}
\usepackage{geometry}
\def\ssection#1{\section{\hbox to \hsize{\large\bf #1\hfill}}}
\def\ssectionstar#1{\section*{\hbox to \hsize{\large\bf #1\hfill}}}
\def\ssubsectionstar#1{\subsection*{\hbox to \hsize{\normalsize\bf #1\hfill}}}
\def\ssubsection#1{\subsection{\hbox to \hsize{\normalsize\bf #1\hfill}}}

\long\def\symbolfootnote[#1]#2{\begingroup%
  \def\thefootnote{\fnsymbol{footnote}}\footnote[#1]{#2}\endgroup%
  \def\footnoterule{\null}}

\definecolor{gray}{gray}{0.85}

\begin{document}

\begin{figure*}[h]
\begin{minipage}[t]{40cm}
\includegraphics[height=35mm]{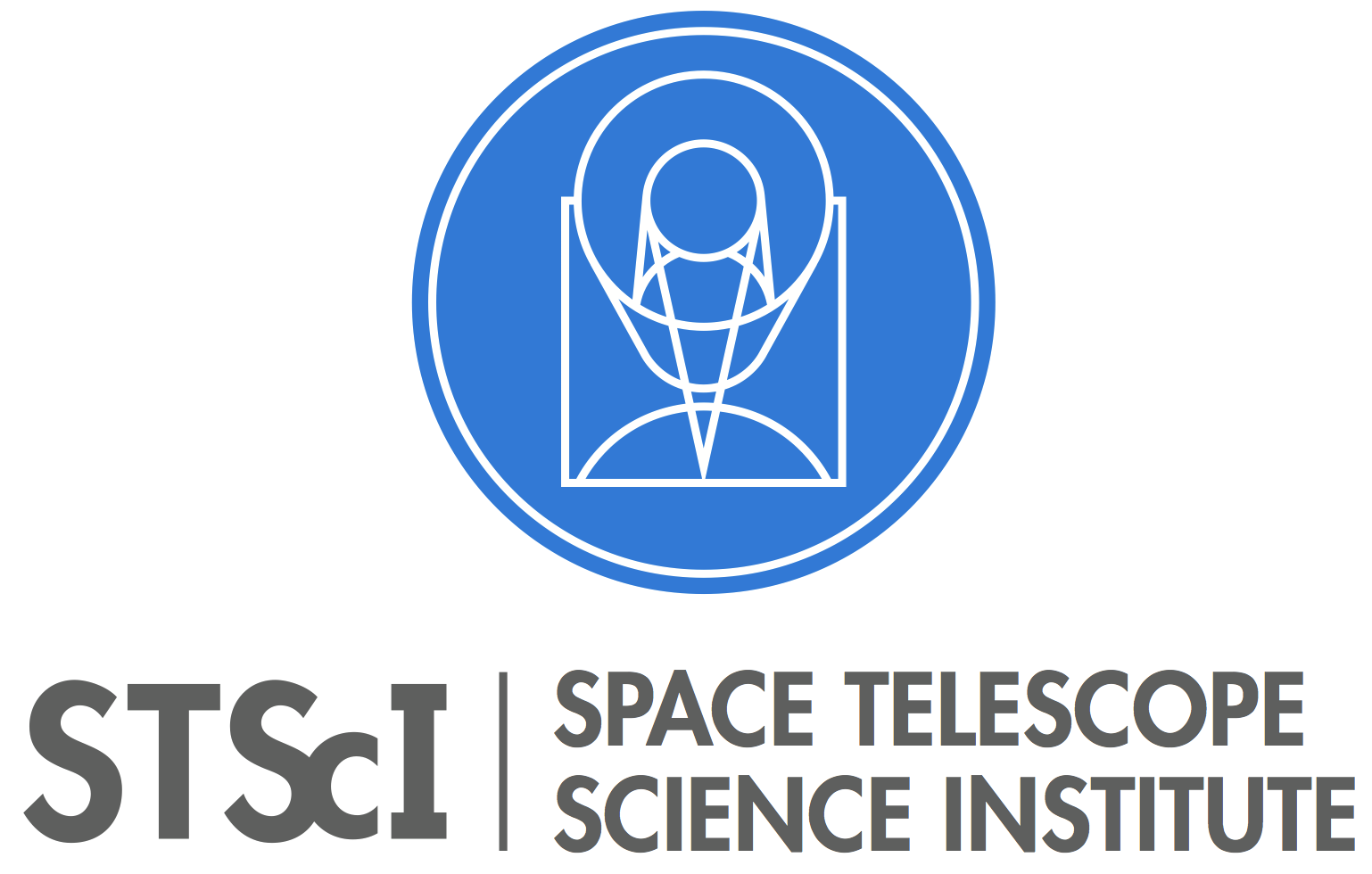}
\end{minipage}
\end{figure*}

\begin{flushright}
\vskip -1.3truecm
{\bf Instrument Science Report WFC3 2019-12}
\end{flushright}

\begin{flushright}
{\huge\bf \hfill Analyzing Eight Years of Transiting Exoplanet Observations Using WFC3's Spatial Scan Monitor}
\rule{145mm}{0.3mm}
\smallskip \\
 \hfill {K.~B. Stevenson \& J. Fowler}\\
 September 30, 2019
 \end{flushright}
 \medskip


\hrule height 1.5pt
\smallskip
\noindent \large{\bf A}\footnotesize{\bf BSTRACT}

\normalsize\noindent{\textit{
HST/WFC3's spatial scan monitor automatically reduces and analyzes time-series data taken in spatial scan mode with the IR grisms.  Here we describe the spatial scan monitor pipeline and present results derived from eight years of transiting exoplanet data.  Our goal is to monitor the quality of the data and make recommendations to users that will enhance future observations.  
We find that a typical observation achieves a white light curve precision that is $1.07\times$ the photon-limit (which is slightly better than expectations) and that the pointing drift is relatively stable during times of normal telescope operations.  We note that observations cannot achieve the optimal precision when the drift along the dispersion direction ($X$ axis) exceeds 15~mas ($\sim$0.11 pixels).  Based on our sample, 77.1\% of observations are ``successful'' ($<15$~mas rms drift), 12.0\% are ``marginal'' (15 -- 135~mas), and 10.8\% of observations have ``failed'' ($>135$~mas or $>1$~pixel), meaning they do not achieve the necessary pointing stability to achieve the optimal spectroscopic precision.  In comparing the observed versus calculated maximum pixel fluence, we find that the J band is a better predictor of fluence than the H band.  Using this information, we derive an updated, empirical relation for scan rate that also accounts for the J-H color of the host star.  We implement this relation and other improvements in version 1.4 of \texttt{PandExo} and version 0.5 of \texttt{ExoCTK}.  Finally, we make recommendations on how to plan future observations with increased precision.
}}

\smallskip
\medskip
\hrule height 1.5pt

\symbolfootnote[0]{Copyright {\copyright} 2019 The Association of
  Universities for Research in Astronomy, Inc. All Rights Reserved.}


\ssection{Introduction}
\normalsize{
Time series exoplanet observations are a common science objective for the Wide Field Camera 3 (WFC3) instrument. By making use of the spatial scan mode \citep{ISR-2012-08} and either of the IR channel's grisms, observations are able to efficiently collect spectroscopic time-series data and achieve higher precision than data previously acquired using WFC3's stare mode \citep{Deming2013}.

The first spatial scan observations took place in 2011 and, as of mid-2019, there are over 250 executed visits.  The majority of these visits are time-series observations of transiting exoplanets, thus providing a treasure trove of data to examine and from which to recommend best practices using this mode.

Below we (1) describe the data reduction software, spatial scan interface, and study population; (2) present our findings from analyzing eight years of time-series data; (3) and provide recommendations to WFC3 users for ways to increase light curve precision and data quality.
}

\ssection{Software \& Analysis}

In this section, we provide a high-level description of the software package and spatial scan interface.  We also list the relevant information on the study population.

\ssubsection{Data Reduction Software}
\label{sec:pipeline}

 Based on analyses first performed by \citet{Stevenson2014a}, we have developed data reduction software to analyze IR grim observations using the spatial scan mode.  As part of the WFC3 Quicklook project \citep{Bourque2017}, we automated this monitor to run daily, identifying and reducing new spatial scan data as they are obtained.
 The outputs from this tool enable us to investigate the performance of time-series observations in spatial scan mode, track data quality over time, and quantify how the quality varies with observational parameters (e.g., pointing drift, fluence, etc.).  Below we provide a high-level description of the pipeline's steps.
\begin{packed_enum}
\item Identify and select WFC3/IR grism spatial scan data then group visits into continuous observations of planets. 
\item Automatically select suitable subarray, background, and spectral extraction regions, which can vary with scan height, position, and grism.
\item Apply a basic flat field correction with no wavelength dependence.
\item Compute difference frames between pairs of non-destructive reads.
\item Perform double outlier rejection (along time axis) of sky background region, automatically selected to be above and below the scanned spectrum.
\item Subtract background region from each difference frame.
\item Apply a rough, integer-pixel pointing drift correction (if drift exceeds 1 pixel).
\item Perform a second double outlier rejection along the time axis, this time incorporating the entire subarray region.
\item Apply a fine, sub-pixel pointing drift/jitter correction.
\item Run optimal spectral extraction \citep{Horne1986}.
\item Shift 1D spectra (along dispersion axis) to align them in pixel space.
\item Compute band-integrated (white) flux by summing values over all non-destructive reads in a given frame.
\item Use the Divide White technique \citep{Stevenson2014a} to remove wavelength-independent systematics and compute spectroscopic light curves.
\end{packed_enum}
The data reduction software makes use of the IMA files rather than the FLT files, as the former yield more robust results by differencing pairs of non-destructive reads.  The reduced data consist of band-integrated light curves (flux vs.~time) with auxiliary information relating to telescope drift/jitter, spectroscopic light curve precision, etc.  The information is stored using Python's object serialization (pickle) format. 

\ssubsection{Spatial Scan Interface}

The spatial scan interface allows us to identify and select the subset of WFC3/IR grism data that use the spatial scan mode.  The WFC3 Quicklook database enables us to quickly identify a subset of FITS files based on header keywords alone, rather than having to open each file to access the required information.  Using a convenience package for this database, \texttt{pyql}, we select the IR grism spatial scan data by using the SCAN\_TYP  == `C' or `D' header keyword from the FITS images. 
This query results in a complete list of every potential observation; however, some exoplanet observations are longer than a single visit, and there may be several such observations in a single program. To appropriately match only a continuous visit, we sought breaks between the starts of subsequent visits that were greater than three hours and thirty minutes (corresponding to three or more {\hst} orbits).
We compared the EXPSTART key from each file, again using \texttt{pyql}, to query for and sort data. 

After passing the data into the reduction pipeline (described in Section \ref{sec:pipeline}), the outputs are saved and displayed on the WFC3 Quicklook website, as part of ongoing monitoring efforts from the WFC3 team. 

\ssubsection{Study Population}

Because the code is fully automated, it has the benefit of yielding uniform results with no human interaction.  However, at times it may yield unexpected results when data are acquired using non-standard observing techniques or when an unexpected event occurs during the observation.  We attempt to account for the former by handling exceptions within the code, and discard the latter.  For the purposes of this statistical study, we consider 166 visits.  Table \ref{tab:params} provides information about our study population as a whole, while Table \ref{tab:obs} at the end of this document provides more detailed information about each target.

\begin{table}[t]
    \makebox[\textwidth][c]{
    \begin{tabular}{|r|l|}
    \hline
    Dates               & 2012.3 -- 2019.5  \\
    J-Band Magnitudes   & 6.07 -- 12.91     \\
    Exposure Times      & 5.97 -- 313.12 sec\\
    Frames per {\hst} orbit & 43 -- 7       \\
    Scan Rate           & 2.0 -- 0.015 arcsec/sec \\
    Forward Mode        & 66 visits         \\
    Round Trip Mode     & 100 visits        \\
    \hline
    \end{tabular}
    }
    \captionsetup{justification=centering}
    \caption{\label{tab:params}
    \textsl{Study population parameters.  The data span a large range of parameters and observing modes.}}
\end{table}


\ssection{Results}
\normalsize{

\ssubsection{Observation Success Rate}

Our first goal in looking at this broad collection of data is to understand the success rate of spatial scan observations as a whole.  Here we define ``success'' as having achieved an rms drift of $\leq$15~mas along the dispersion direction ($X$ axis).
As discussed in Section \ref{sec:pointing}, the success of an observation does not strongly depend on drift along the spatial direction ($Y$ axis).  We classify an observation as having ``failed'' when the rms drift in $X$ is $>$135~mas ($>$1 pixel).  We classify drifts between 15 and 135~mas, where drawing meaningful conclusions from the data is visit dependent, as ``marginal'' observations.  We discuss these classifications and their impact on the data in later sections.

Figure \ref{fig:hist} demonstrates that, over the lifetime of WFC3 spatial scan observations, 77.1\% of visits are successful.  Correspondingly, 10.8\% of visits have failed.  Within the subset of failed visits we look for trends such as higher scan rates, localized observation dates, and common target positions.  We identify a relatively large number of failures in 2018, corresponding to a time when {\hst} experienced increased gyro bias levels and more frequent guide star acquisition failures.
Otherwise, we find no statistically significant deviations relative to the successful visits (see Figure \ref{fig:sf}). Common reasons for a failed visit are a guide star acquisition failure, observing in gyro mode [i.e., during South Atlantic Anomaly (SAA) crossings], and use of a single guide star.  Typically, only the first reason is eligible for a repeat observation.

\begin{figure}[tp!]
\begin{center}
\makebox[\textwidth][c]{\includegraphics[width=0.85\linewidth]{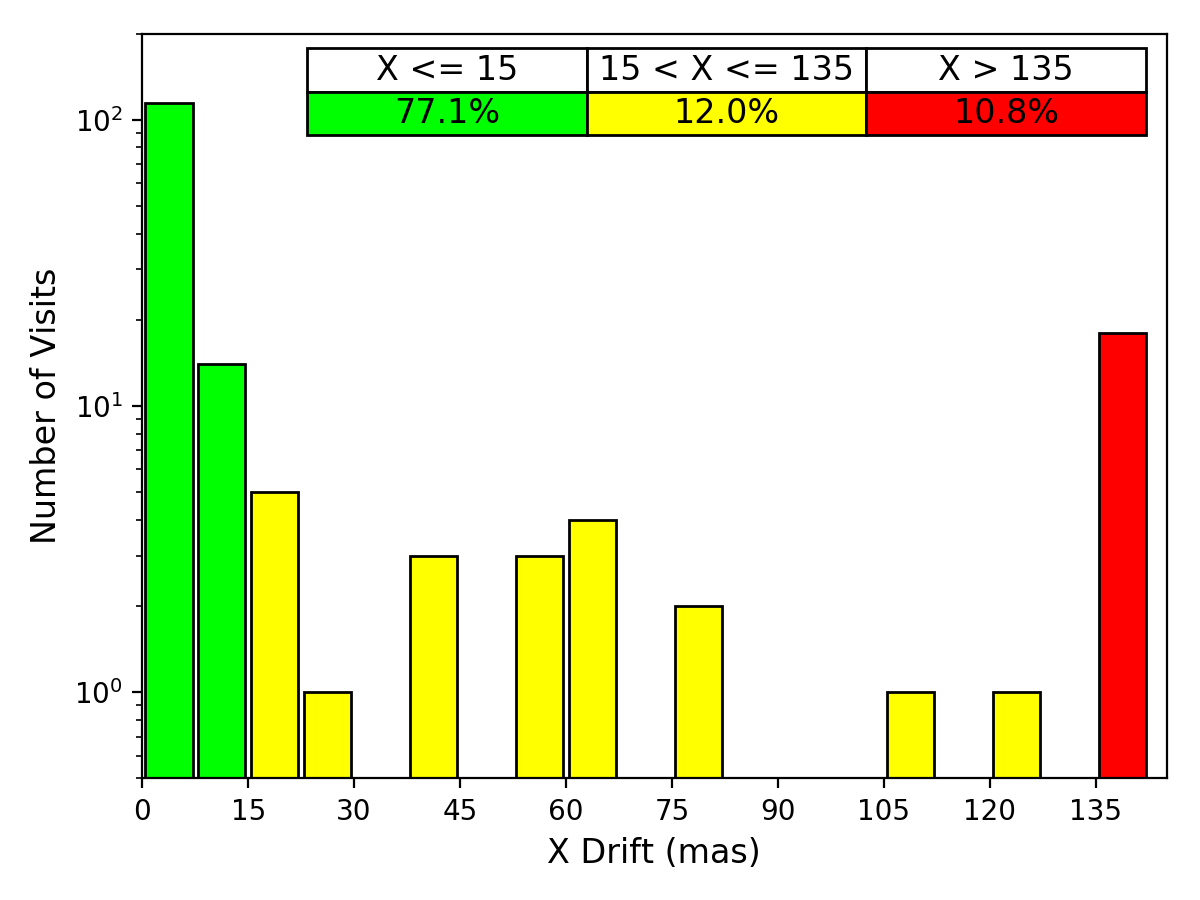}}\vspace{-1em}
\caption{\label{fig:hist}
\textsl{Histogram of measured spectrum drift along the $X$ axis for all spatial scan observations in our study.  The right-most bin contains all observations with drift $>$135~mas ($>$1 pixel).  Our analyses indicate that 77.1\% of visits are successful, whereas 10.8\% of visits have failed.
}}
\end{center}

\begin{center}
\makebox[\textwidth][c]{\includegraphics[width=1.0\linewidth]{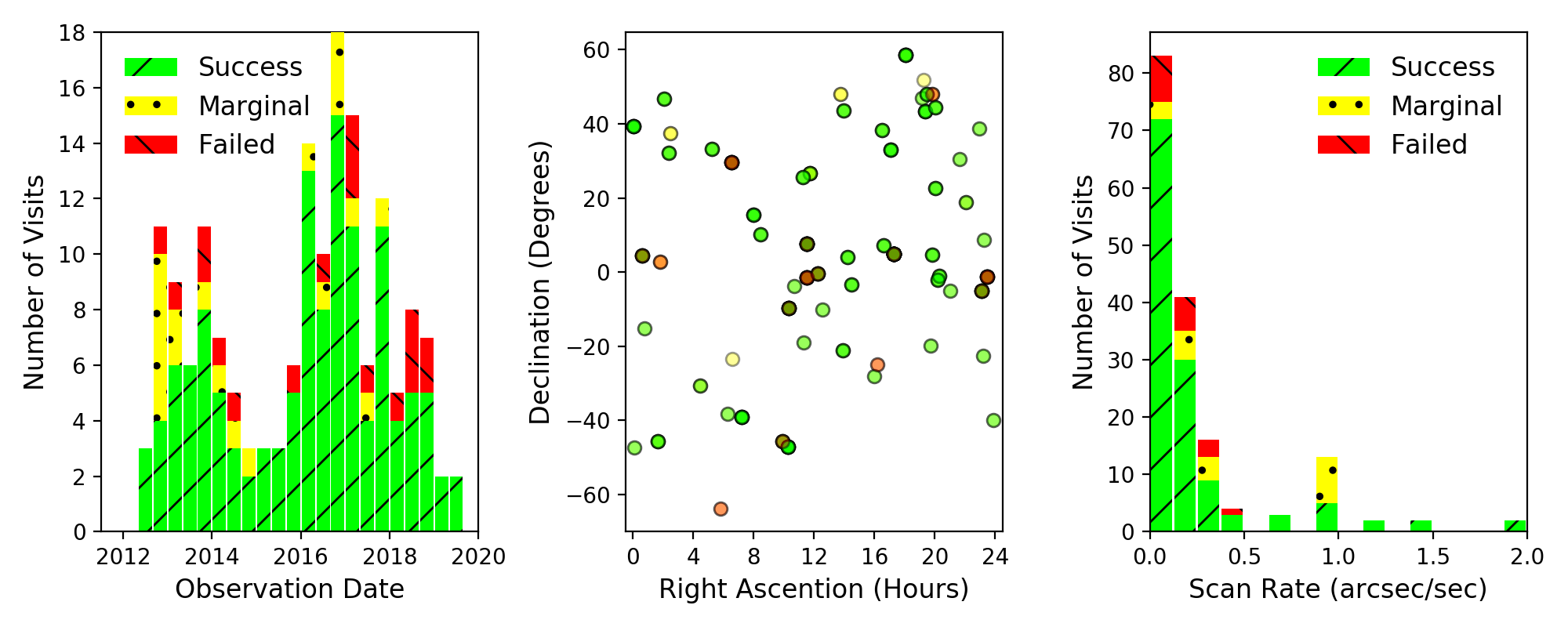}}\vspace{-1em}
\caption{\label{fig:sf}
\textsl{Observation success rate shown as functions of the observation date, position, and scan rate.  There is no obvious trend outside of the increased failure rate in 2018 when {\hst} is known to have experienced gyro issues.
}}
\end{center}
\end{figure}

\ssubsection{Effects of Pointing Drift}
\label{sec:pointing}

Figure \ref{fig:drift} displays the measured drift along the $X$ and $Y$ axes for all successful visits.  The data suggest that the drift may have been slightly elevated in 2018 when {\hst} experienced issues with its gyros.  Otherwise, the rolling median rms drift shows no significant or lasting pointing degradation.  The median drift of the successful visits along $X$ and $Y$ are $5.1^{+1.9}_{-1.5}$ and $4.0^{+4.3}_{-1.4}$~mas, respectively. There is also no statistically significant difference in drift between observations that scan only in the FORWARD direction versus those that use the ROUND\_TRIP mode.

\begin{figure}[t]
\begin{center}
\makebox[\textwidth][c]{\includegraphics[width=0.8\linewidth]{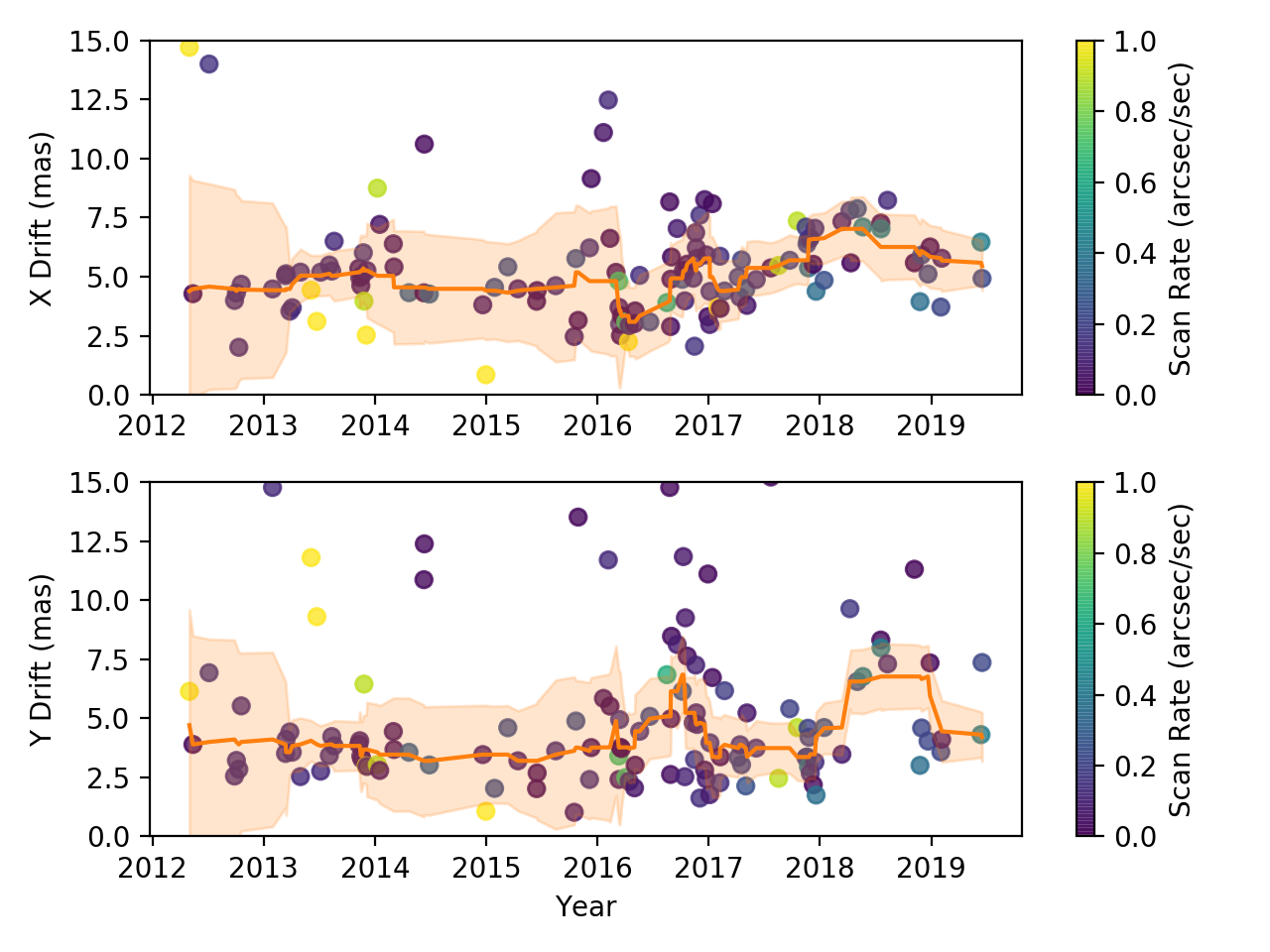}}\vspace{-1em}
\caption{\label{fig:drift}
\textsl{Measured pointing drift (in milliarcseconds) over $>7$ years of spatial scan observations for successful observations (where the drift along $X$ is $<15$ mas).  The pointing drift within a visit has changed very little over the years ($5.0 \pm 2.3$, $4.0 \pm 3.4$ mas).  The orange line and $1\sigma$ regions are computed using a rolling median over 15 neighboring visits.  Observations with a higher scan rate do not exhibit a systematically larger drift.
}}
\end{center}
\end{figure}

Next we consider the effects of pointing drift on the quality of the data.  Figure \ref{fig:scatter} shows the scatter in the normalized spectroscopic light curves as functions of the measured drifts along the $X$ and $Y$ axes.  We note that the best precision achieved when the drift along the $X$ axis exceeds 15~mas is 460~ppm.  This can be compared to 153~ppm precision for successful observations ($X$ drift $<$ 15~mas).  As seen in the right panel of Figure~\ref{fig:scatter}, this correlation does not hold for drift along the $Y$ axis.  This makes sense since {\hst} scans in the spatial direction ($Y$ axis) and the flux is expected to be constant along that direction.  

Based on this line of evidence, we conclude that a reasonable delineator for success is an rms drift of 15~mas along the dispersion direction.  
Delineating between a marginal and failed observation requires a more qualitative argument because there is no abrupt transition.  The low precision light curves with large drifts are likely the result of residual, uncorrected flat field effects.  Thus, when the drift is $<1$ pixel, this systematic can often be accounted for using relatively simple models.  With larger drifts, we have been unsuccessful in our attempts to adequately correct this systematic due to its complicated nature and significant amplitude relative to the signals seen in transmission spectroscopy.

\begin{figure}[t]
\begin{center}
\makebox[\textwidth][c]{\includegraphics[width=0.8\linewidth]{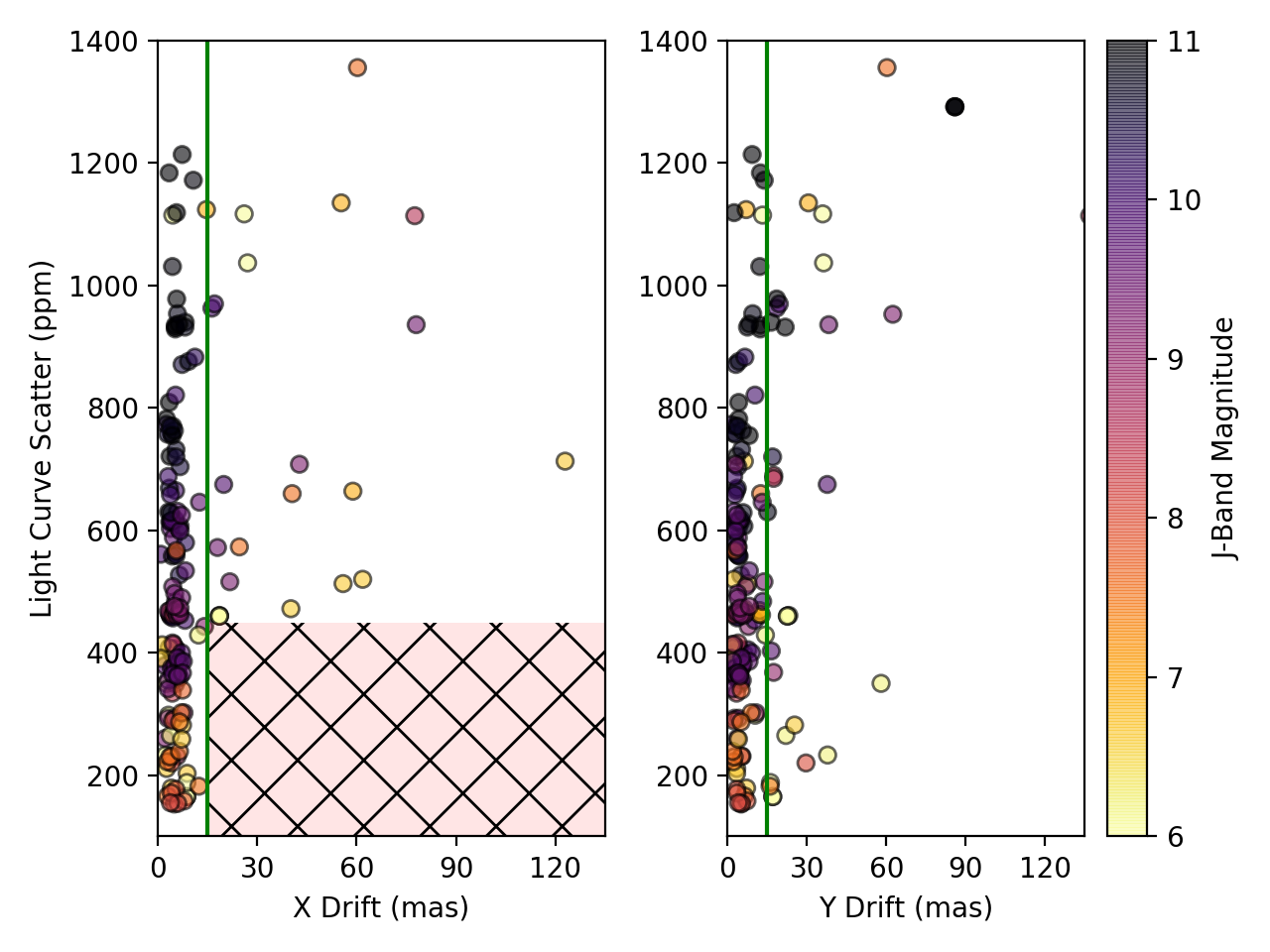}}\vspace{-1em}
\caption{\label{fig:scatter}
\textsl{Measured scatter in the normalized spectroscopic light curves as functions of the measured drift along the $X$ and $Y$ axes.  The red cross-hatched region is devoid of observations, leading us to conclude that observations cannot achieve the optimal precision when the drift along the dispersion direction ($X$ axis) exceeds 15~mas ($\sim$0.11 pixels).
}}
\end{center}
\end{figure}


\ssubsection{Observation Planning}

Based on the successful observational data, we develop new empirical equations to predict the maximum pixel fluence, $F$, and desired scan rate, $S$, for the G141 grism.  As seen in Figure~\ref{fig:fluence}, the equation previously used by \texttt{PandExo} \citep[Equation~\ref{eqn:h},][]{Batalha2017} does not adequately predict the observed fluence for many of the observations.  This is because Equation~\ref{eqn:h} relies on the H-band magnitude, which is redward of G141's measured peak flux and does not account for the stellar type (see Figure \ref{fig:spectra}).  Equation~\ref{eqn:j} yields a better fit by using the J-band magnitude;  however, even with J-band there is still a small color dependence.  To account for the stellar type, we adopt Equation~\ref{eqn:color} as our best-fit model.  The standard deviations of the residuals when applying all three equations to the measured maximum pixel fluence level are: 5149, 3310, and 2738 e\sp{-}/pix, respectively.  Thus, Equation~\ref{eqn:color} reduces the scatter by a factor of 1.9 relative to Equation~\ref{eqn:h}.
\begin{equation}\label{eqn:h}
    F_{H} = \frac{13.2}{S}10^{-0.4(h_{mag}-15)} \mathrm{~e\sp{-}/pix}
\end{equation}
\begin{equation}\label{eqn:j}
    F_{J} = \frac{2365}{S}10^{-0.4(j_{mag}-9.75)} \mathrm{~e\sp{-}/pix}
\end{equation}
\begin{equation}\label{eqn:color}
    {\bf F_{J+Color} = \frac{2491}{S}10^{-0.4(j_{mag}-9.75)} - 
              \frac{ 161}{S}10^{-0.4(j_{mag}-h_{mag})} \mathrm{~e\sp{-}/pix}}
\end{equation}

\noindent Rearranging Equation~\ref{eqn:color} to compute the scan rate as a function of the desired maximum pixel fluence (in e\sp{-}/pix) is trivial:
\begin{equation}\label{eqn:scanrate}
    {\bf S = \frac{2491}{F}10^{-0.4(j_{mag}-9.75)} - 
             \frac{ 161}{F}10^{-0.4(j_{mag}-h_{mag})} \mathrm{~arcsec/sec}}.
\end{equation}
\noindent Typically, $F = 30k$ e\sp{-}/pix; however, as discussed in Section \ref{sec:precision}, there may be good reasons to choose higher fluence values for relatively bright targets.  If the H-band magnitude is unknown, users can assume no color dependence and negate the second term of Equation~\ref{eqn:scanrate}.  To first order, the G102 scan rate is still 80\% of the G141 scan rate.  

\begin{figure}[tp!]
\begin{center}
\makebox[\textwidth][c]{\includegraphics[width=1.0\linewidth]{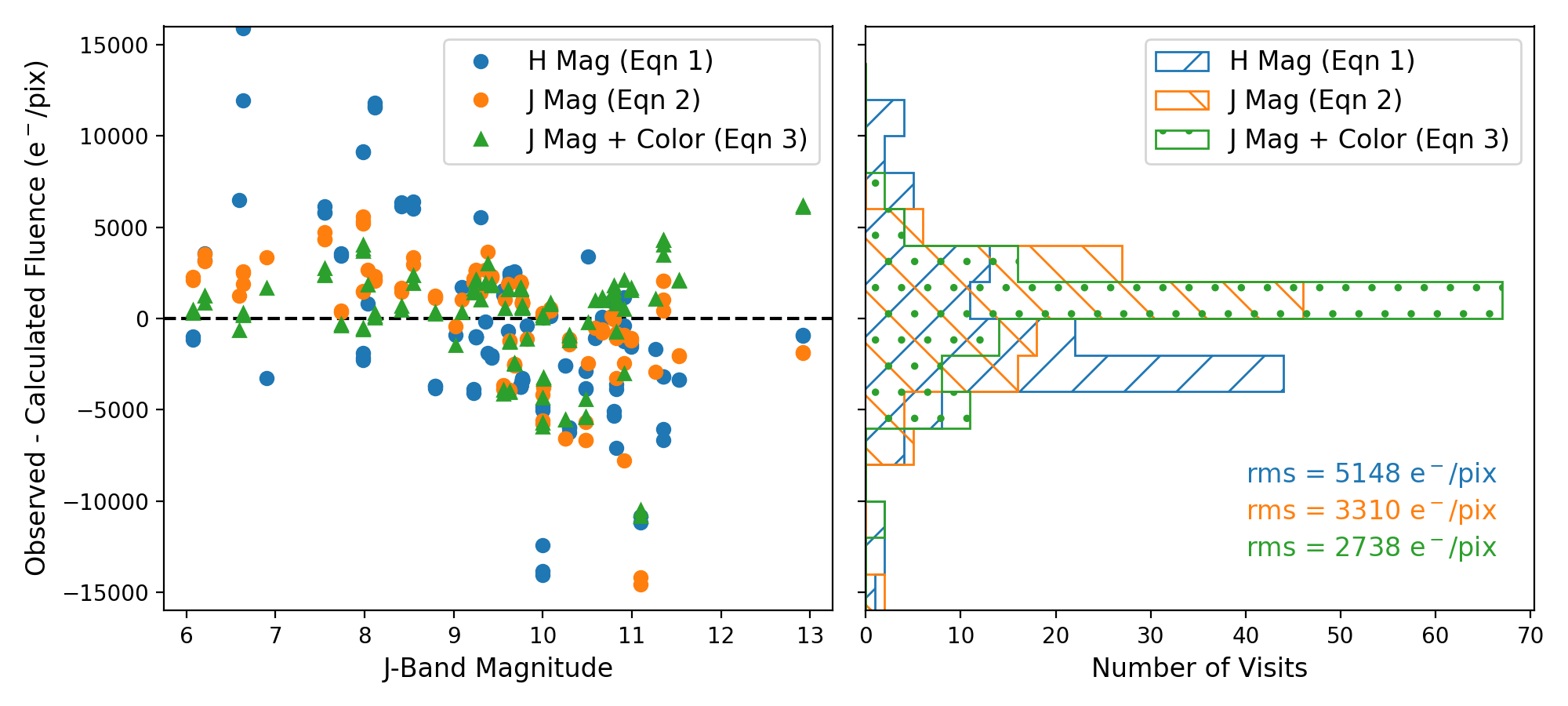}}
\caption{\label{fig:fluence}
\textsl{Residual fluence levels for successful G141 observations.  Using J-band magnitudes and a color correction (Equation~\ref{eqn:color}) yields the best fit to the observed fluence levels.
}}
\end{center}
\vspace{1em}
\begin{center}
\makebox[\textwidth][c]{\includegraphics[width=1.0\linewidth]{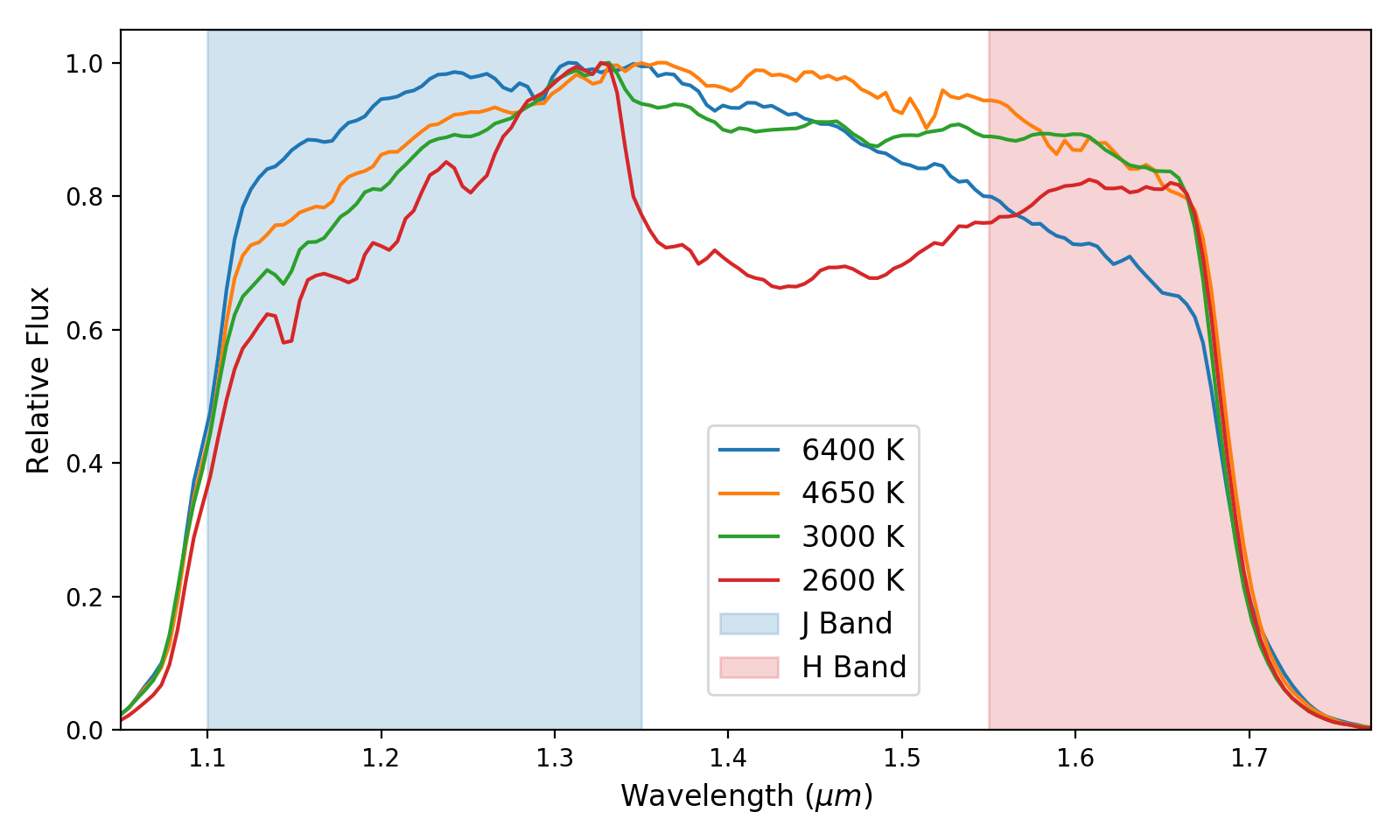}}
\caption{\label{fig:spectra}
\textsl{WFC3/G141 stellar spectra at various temperatures.  A star's J-band magnitude better represents its maximum pixel fluence, which peaks near 1.3 {\microns} for the range of stellar types shown here.
}}
\end{center}
\end{figure}

\ssubsection{Light Curve Precision}
\label{sec:precision}

First, we investigate how well the WFC3 G141 observations perform with respect to theoretical predictions.  Figure \ref{fig:precision} depicts the band-integrated (white) light curve precisions and compares them to the anticipated photon-limited precisions.  The measured values closely follow, but are consistently above, the theoretical predictions because other noise sources are not considered.  Our best-fit solution is $1.07\times$ the photon-limited precision curve.  We recommend that users use this multiplier when computing rough signal-to-noise estimates for their targets.

\begin{figure}[tp]
\begin{center}
\mbox{\includegraphics[width=0.75\linewidth]{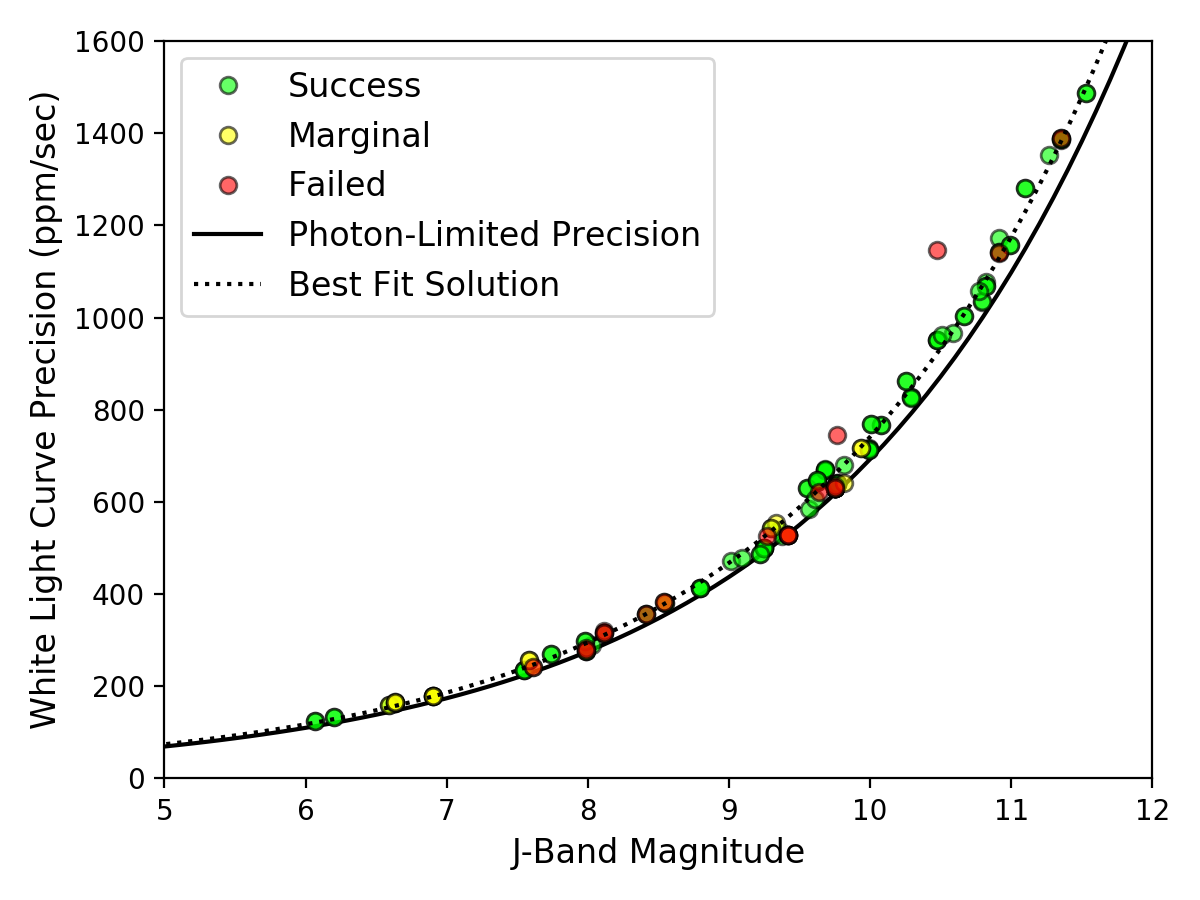}}\vspace{-1em}
\caption{\label{fig:precision}
\textsl{Comparing measured white light curve precisions to theoretical predictions.  The best-fit solution is $1.07\times$ the photon-limited precision, which is smaller than the value adopted by \texttt{PandExo\_HST} (1.14). To achieve a uniform comparison, the values are plotted per second of exposure time.
}}
\end{center}
\end{figure}

Next, we compute the predicted spectroscopic light curve precision using \texttt{PandExo\_HST} \citep{Batalha2017}.  By plotting the precision (relative to the default configuration; 150 s, 30k e$^-$/pix, Round Trip) versus magnitude, Figure \ref{fig:relativerms} demonstrates that targets brighter than J = 8 can typically achieve a higher precision by increasing the per-pixel fluence (compare solid lines, negative values represent a higher precision relative to the default configuration).  Specifically, one must decrease the scan rate and increase the number of samples (i.e. non-destructive reads) per frame, thus resulting in an increase in the per-pixel fluence over a similar number of pixels and an overall higher precision.  
Fainter targets (J $>$ 8) may benefit from using longer exposure times (see dotted line in Figure \ref{fig:relativerms}); however, care must be taken to not overly degrade the time resolution of the data.
Figure \ref{fig:relativerms} also depicts the improvement in precision when using the ROUND\_TRIP spatial scan mode instead of the FORWARD mode (dashed line).

\begin{figure}[tp]
\begin{center}
\mbox{\includegraphics[width=0.9\linewidth]{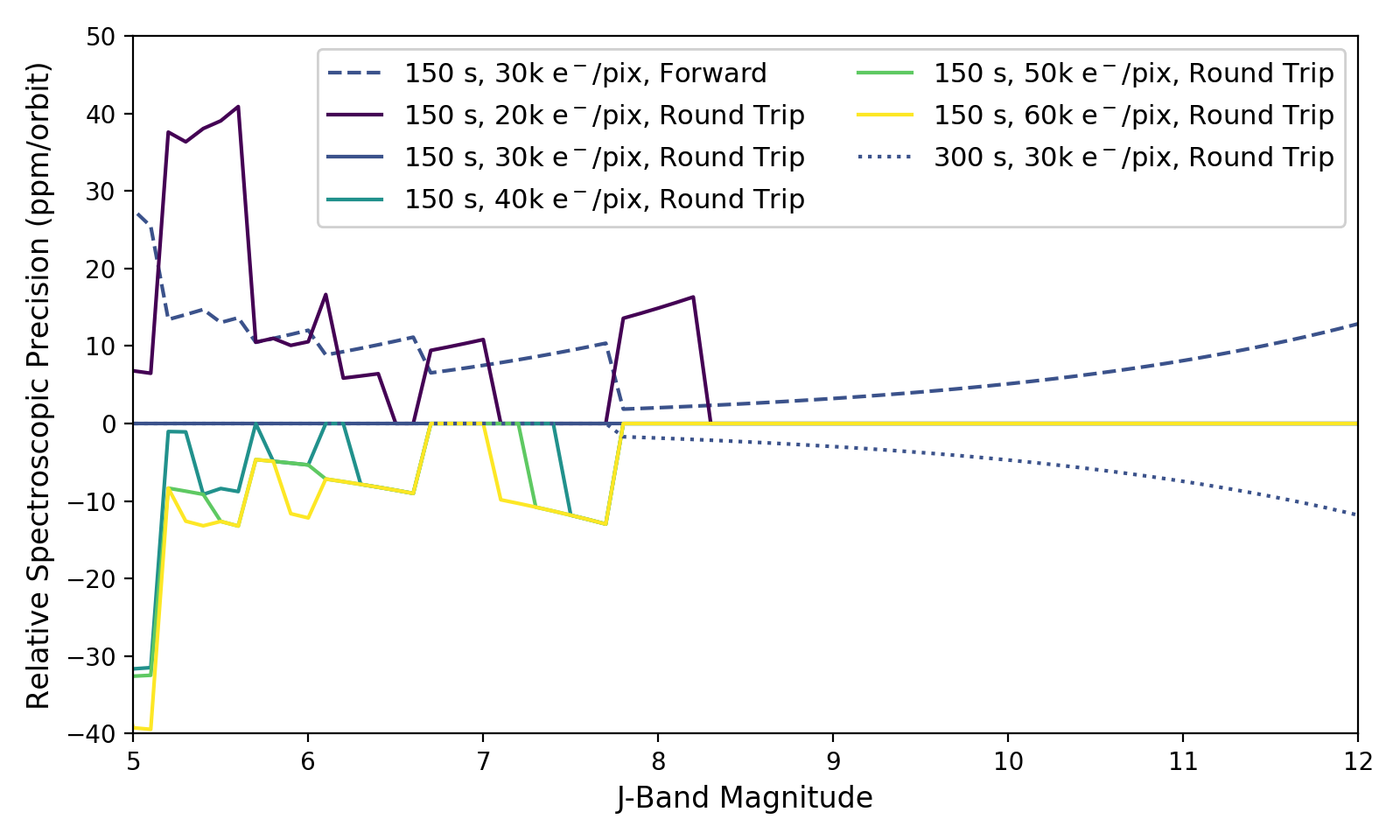}}\vspace{-1em}
\caption{\label{fig:relativerms}
\textsl{Relative spectroscopic light curve precisions per {\hst} orbit assuming 16, 7-pixel-wide channels.  These predictions use \texttt{PandExo\_HST}'s optimal observing strategy and consider two maximum exposure times (150 \& 300 seconds), five fluence levels (20k -- 60k e$^-$/pix), and two spatial scan modes (Forward and Round Trip).  Predictions are relative to \texttt{PandExo\_HST}'s default configuration (150 s, 30k e$^-$/pix, Round Trip).
}}
\mbox{\includegraphics[width=1.\linewidth]{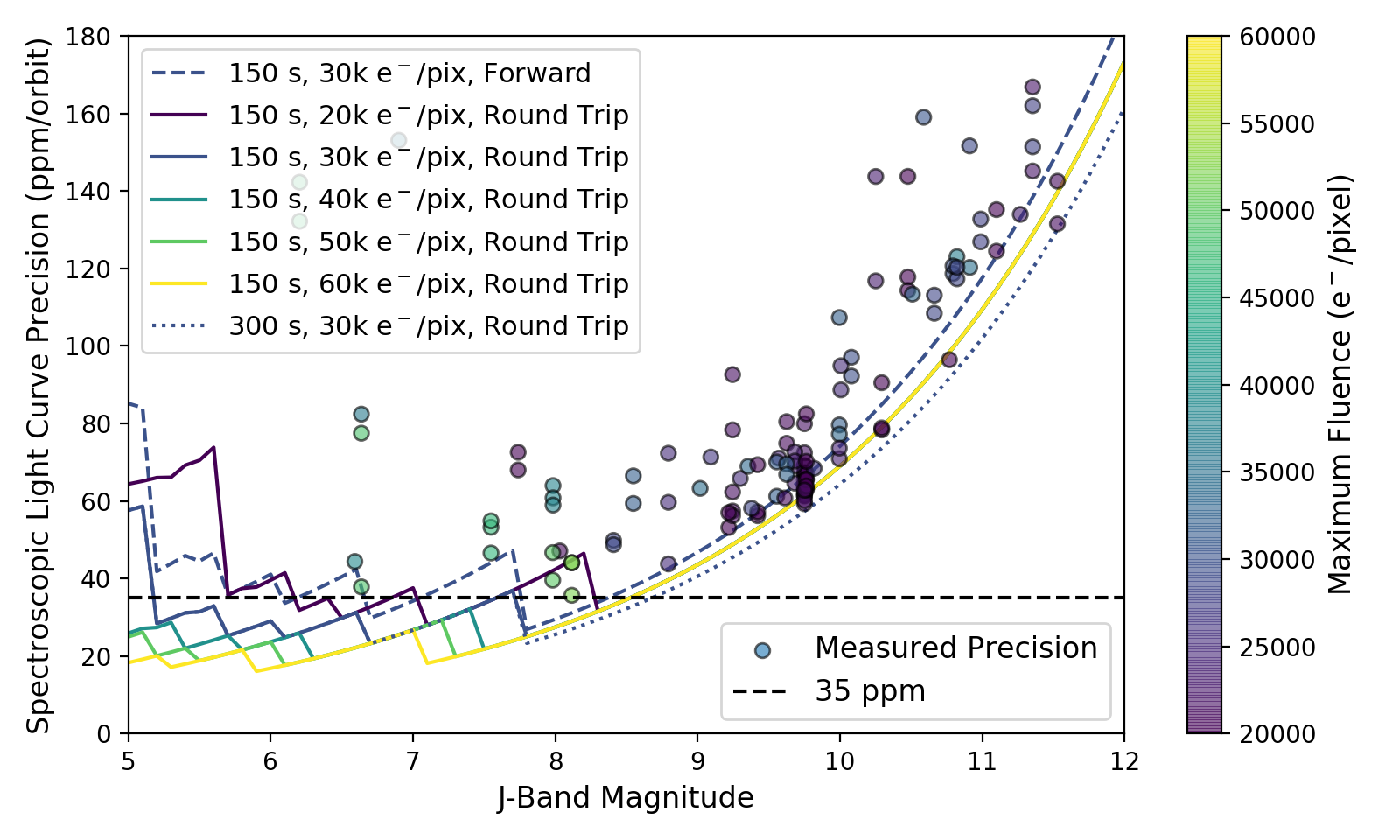}}\vspace{-1em}
\caption{\label{fig:pandexo}
\textsl{Measured and predicted spectroscopic light curve precisions per {\hst} orbit.  See Figure \ref{fig:relativerms}'s caption for details.  Regardless of magnitude, all measured values exceed 35~ppm/orbit (dashed line) at $R\sim40$.
}}
\end{center}
\end{figure}

Finally, we compute the measured spectroscopic light curve precision by use of the Divide White method \citep{Stevenson2014a}, which removes wavelength-independent systematics.  We do not take the additional step of removing wavelength-dependent variations in the transit/eclipse depths as these values are small compared to the point-to-point scatter.  Each of the 16 spectroscopic bins is 7 pixels in width.  The final spectroscopic precision is a median of the 16 values recorded from each observation.

In Figure \ref{fig:pandexo}, the measured spectroscopic precision is typically within 50~ppm of predictions from \texttt{PandExo\_HST} \citep{Batalha2017}.  Some programs did not adopt the most efficient observing strategy and others experienced wavelength-dependent systematics, which weren't accounted for in our automated data reduction and analysis pipeline.  Both scenarios lead to a degradation in measured precision relative to the idealized predictions from \texttt{PandExo\_HST}.  The precision per {\hst} orbit levels off once the detector read plus reset time is significantly greater than the exposure time.  This fact is important when attempting to identify the highest signal-to-noise targets.  For example, observations of HD~189733 (J = 6.1) are likely to yield a similar precision to those of WASP-18 (J = 8.4), assuming the same maximum pixel fluence.  For the brightest targets, there appears to be an increasing departure from theoretical predictions, potentially hinting at the presence of a systematic noise floor of at most 21~ppm for a resolving power of $\sim$40.  More work is needed to further explore and better quantify this tentative result, which is beyond the scope of this ISR.

Although Figure \ref{fig:relativerms} predicts higher precisions for the ROUND\_TRIP mode, we investigate whether this mode truly performs better than the FORWARD mode.  Looking specifically at the successful observations of GJ~1214, we see that all nine visits in ROUND\_TRIP mode achieve a higher precision per {\hst} orbit than any of the five visits using FORWARD mode.  This is because WFC3 can acquire more scans per {\hst} orbit using ROUND\_TRIP mode, which has smaller overheads compared to the FORWARD mode.  The median improvement between the two modes is 6.4~ppm per spectroscopic channel, which is consistent with \texttt{PandExo\_HST}'s prediction of 4.6~ppm.  More broadly, this trend repeats across other observations; therefore, we recommend that all future time-series observations use the ROUND\_TRIP mode.

}


\ssection{Summary of Recommendations}
\normalsize{

Using WFC3's spatial scan monitor, we have analyzed eight years of time-series data and report on the success rate of these observations.
Based on these findings, we make several recommendations that will help maximize the efficiency of spatial scan, time-series observations and enhance users' understanding of its limits.

\squishlist
\item To minimize pointing drift, observers should always use the Fine Guidance Sensors (FGS mode) with multiple guide stars and avoid SAA crossings at the beginning of {\hst} orbits.
\item For the most precise results, time-series observations should always implement the ROUND\_TRIP spatial scan mode.
\item Time-series observations of brighter targets (J $< 8$) may adopt a max pixel fluence greater than 30k e\sp{-}/pix to achieve a higher precision. 
\item Time-series observations of fainter targets (J $> 8$) may adopt exposure times longer than 150 seconds to achieve a higher precision.
\item Observers should assume a best-case precision of 35~ppm per {\hst} orbit at $R\sim40$ (16~channels).
\squishend
}

\ssectionstar{Acknowledgements}

We would like to thank Marc Rafelski for his thorough review of this ISR.  We would also like to thank the WFC3 team for their useful insights and discussions on this program.


{
\bibliography{ms}

\begin{thebibliography}{}
\expandafter\ifx\csname natexlab\endcsname\relax\def\natexlab#1{#1}\fi

\bibitem[{{Batalha} {et~al.}(2017){Batalha}, {Mandell}, {Pontoppidan},
  {Stevenson}, {Lewis}, {Kalirai}, {Earl}, {Greene}, {Albert}, \&
  {Nielsen}}]{Batalha2017}
{Batalha}, N.~E., {et~al.} 2017, \pasp, 129, 064501

\bibitem[{{Bourque} {et~al.}(2017){Bourque}, {Bajaj}, {Bowers}, {Dulude},
  {Durbin}, {Gosmeyer}, {Gunning}, {Khandrika}, {Martlin}, {Sunnquist}, \&
  {Viana}}]{Bourque2017}
{Bourque}, M., {et~al.} 2017, in IAU Symposium, Vol. 325, Astroinformatics, ed.
  M.~{Brescia}, S.~G. {Djorgovski}, E.~D. {Feigelson}, G.~{Longo}, \&
  S.~{Cavuoti}, 397--400

\bibitem[{{Deming} {et~al.}(2013){Deming}, {Wilkins}, {McCullough}, {Burrows},
  {Fortney}, {Agol}, {Dobbs-Dixon}, {Madhusudhan}, {Crouzet}, {Desert},
  {Gilliland}, {Haynes}, {Knutson}, {Line}, {Magic}, {Mandell}, {Ranjan},
  {Charbonneau}, {Clampin}, {Seager}, \& {Showman}}]{Deming2013}
{Deming}, D., {et~al.} 2013, \apj, 774, 95

\bibitem[{{Horne}(1986)}]{Horne1986}
{Horne}, K. 1986, Publ. Astron. Soc. Pac., 98, 609

\bibitem[{{McCullough} \& {MacKenty}(2012)}]{ISR-2012-08}
{McCullough}, P., \& {MacKenty}, J. 2012, {Considerations for using Spatial
  Scans with WFC3}, Tech. rep.

\bibitem[{{Stevenson} {et~al.}(2014){Stevenson}, {Bean}, {Seifahrt},
  {D{\'e}sert}, {Madhusudhan}, {Bergmann}, {Kreidberg}, \&
  {Homeier}}]{Stevenson2014a}
{Stevenson}, K.~B., {et~al.} 2014, \aj, 147, 161

\end{thebibliography}
}
\clearpage

\setlength{\tabcolsep}{4pt}
\renewcommand{\arraystretch}{0.87}
\begin{center}
\begin{ThreePartTable}
\begin{longtable}{rlrrrlrr}
\captionsetup{justification=centering}
\caption{\label{tab:obs} Summary of Observations.}\\
\hline
Obs.    & Target& Fluence\sp{a} & X rms & Y rms & Scan Type    & Scan Rate  & Precision\sp{b}     \\
Date    &       & (e\sp{-}/pix) & (mas) & (mas) &              & ("/sec)    & (ppm/orbit)   \\
\hline
\endfirsthead
\multicolumn{8}{c}%
{\tablename\ \thetable\ -- \textit{Continued from previous page}} \\
\hline
Obs.    & Target& Fluence\sp{a} & X rms & Y rms & Scan Type    & Scan Rate  & Precision\sp{b}     \\
Date    &       & (e\sp{-}/pix) & (mas) & (mas) &              & ("/sec)    & (ppm/orbit)   \\
\hline
\endhead
\hline \multicolumn{8}{r}{\textit{Continued on next page}} \\
\endfoot
\hline
\endlastfoot
 2012.33 & GJ 436       & 36322 &  14.7 &   6.1 & Forward    & 0.99  &   153 \\
 2012.36 & WASP-31      & 34904 &   4.3 &   3.9 & Forward    & 0.019 &   120 \\
 2012.51 & HAT-P-1      & 44220 &  14   &   6.9 & Forward    & 0.15  &    57 \\
 2012.73 & HD 209458    & 49390 &  40.1 &   5.6 & Forward    & 0.9   &    43 \\
 2012.74 & GJ 1214      & 21643 &   4   &   2.6 & Forward    & 0.12  &    69 \\
 2012.76 & GJ 1214      & 21636 &   4.3 &   3.2 & Forward    & 0.12  &    80 \\
 2012.78 & GJ 1214      & 21686 &   2   &   2.8 & Forward    & 0.12  &    69 \\
 2012.8  & HAT-P-11     & 49974 &  60.2 &  54   & Forward    & 0.37  &    39 \\
 2012.8  & GJ 1214      & 21628 &   4.7 &   5.5 & Forward    & 0.12  &    67 \\
 2012.82 & GJ 436       & 36232 &  55.3 &  27.4 & Forward    & 0.99  &    67 \\
 2012.9  & WASP-33      & 69169 &  40.5 &  11.2 & Forward    & 0.25  &    43 \\
 2012.91 & GJ 436       & 36205 &  58.8 &  28.1 & Forward    & 0.99  &    63 \\
 2012.94 & GJ 436       & 35955 &  63.8 &  31   & Forward    & 0.99  &    65 \\
 2012.96 & HAT-P-11     & 50096 & 398   & 281.7 & Forward    & 0.37  &    80 \\
 2013.01 & GJ 436       & 36026 &  60.4 &  25.8 & Forward    & 0.99  &    65 \\
 2013.04 & WASP-33      & 69200 &  24.5 &   3.5 & Forward    & 0.25  &    46 \\
 2013.08 & GJ 1214      & 21685 &   4.5 &  14.8 & Forward    & 0.12  &    72 \\
 2013.2  & GJ 1214      & 21656 &   5   &   3.5 & Round\_Trip & 0.12  &    61 \\
 2013.2  & GJ 1214      & 21648 &   5.1 &   4.1 & Round\_Trip & 0.12  &    62 \\
 2013.23 & GJ 1214      & 21711 &   3.5 &   4.4 & Round\_Trip & 0.12  &    61 \\
 2013.26 & GJ 1214      & 21666 &   3.7 &   3.6 & Round\_Trip & 0.12  &    67 \\
 2013.28 & GJ 1214      & 21633 & 169.4 & 435   & Round\_Trip & 0.12  &    84 \\
 2013.33 & GJ 1214      & 21608 &   5.2 &   2.5 & Round\_Trip & 0.12  &    59 \\
 2013.43 & HD 189733    & 37170 &   4.4 &  11.8 & Round\_Trip & 2     &   983 \\
 2013.48 & HD 189733    & 37338 &   3.1 &   9.3 & Round\_Trip & 2     &   209 \\
 2013.51 & GJ 1214      & 21618 &   5.2 &   2.8 & Round\_Trip & 0.12  &    63 \\
 2013.59 & GJ 1214      & 21642 &   5.5 &   3.4 & Round\_Trip & 0.12  &    64 \\
 2013.61 & GJ 1214      & 21641 &   5.2 &   4.2 & Round\_Trip & 0.12  &    60 \\
 2013.63 & GJ 1214      & 21658 &   6.5 &   3.8 & Round\_Trip & 0.12  &    65 \\
 2013.82 & HD 165459    & 49029 & 123   &   5.7 & Round\_Trip & 0.9   &   155 \\
 2013.85 & WASP-43      &     7 & nan   & 614.9 & Round\_Trip & 0.08  &   nan \\
 2013.86 & WASP-43      & 31973 &   5.4 &   3.9 & Round\_Trip & 0.05  &    80 \\
 2013.86 & WASP-43      & 23733 &   5   &   4.1 & Round\_Trip & 0.08  &    71 \\
 2013.87 & WASP-43      & 32178 &   4.6 &   3.4 & Round\_Trip & 0.05  &   107 \\
 2013.88 & WASP-43      & 33580 &   5   &   3.3 & Round\_Trip & 0.05  &    77 \\
 2013.9  & HAT-P-17     & 34226 &   6   &  15.6 & Forward    & 0.134 &    63 \\
 2013.9  & HD 165459    & 48718 &   4   &   6.5 & Round\_Trip & 0.9   &    77 \\
 2013.92 & HD 165459    & 36548 &   2.5 &   3   & Round\_Trip & 1.2   &    82 \\
 2013.93 & WASP-43      & 23883 &   5.3 &   3   & Round\_Trip & 0.08  &    74 \\
 2013.96 & WASP-12      &     7 & nan   & nan   & Forward    & 0.05  &   nan \\
 2014    & WASP-12      & 14261 & 584.7 & 288.9 & Round\_Trip & 0.05  & 16725 \\
 2014.02 & HD 165459    & 48789 &   8.8 &   3   & Round\_Trip & 0.9   &    38 \\
 2014.04 & WASP-12      & 18532 &   7.2 &   2.8 & Round\_Trip & 0.05  &   144 \\
 2014.16 & WASP-12      & 17561 &   6.4 &   4.4 & Round\_Trip & 0.05  &   114 \\
 2014.17 & WASP-12      & 17552 &   5.4 &   3.7 & Round\_Trip & 0.05  &   118 \\
 2014.31 & WASP-18      & 28581 &   4.3 &   3.6 & Round\_Trip & 0.3   &    50 \\
 2014.33 & HD 165459    & 48787 &  61.8 &   2.1 & Round\_Trip & 0.9   &   126 \\
 2014.35 & WASP-18      & 28490 & 110.5 & 463.8 & Round\_Trip & 0.3   &   122 \\
 2014.44 & WASP-19      & 30316 &   4.3 &  10.9 & Forward    & 0.026 &   310 \\
 2014.44 & WASP-19      & 28759 &  10.6 &  12.4 & Forward    & 0.026 &   152 \\
 2014.45 & WASP-19      & 29614 & 208.8 & 346.1 & Forward    & 0.026 &   186 \\
 2014.49 & WASP-18      & 28797 &   4.3 &   3   & Round\_Trip & 0.3   &    49 \\
 2014.67 & HD 165459    & 48697 &  55.9 &   6.9 & Round\_Trip & 0.9   &   112 \\
 2014.97 & KEPLER-138   & 19344 &   3.8 &   3.5 & Round\_Trip & 0.07  &    78 \\
 2015    & HD 209458    & 38979 &   0.8 &   1.1 & Round\_Trip & 1.15  &    44 \\
 2015.08 & GJ 3470      & 24989 &   4.5 &   2   & Round\_Trip & 0.24  &    72 \\
 2015.19 & GJ 3470      & 24998 &   5.4 &   4.6 & Round\_Trip & 0.24  &    44 \\
 2015.29 & KEPLER-138   & 19357 &   4.5 &   3.2 & Round\_Trip & 0.07  &    79 \\
 2015.45 & 2MASS J16371 & 12736 &   4   &   2   & Round\_Trip & 0.025 &   124 \\
 2015.46 & 2MASS J16371 & 13097 &   4.4 &   2.7 & Round\_Trip & 0.025 &   135 \\
 2015.62 & LHS 6343     & 24293 &   4.6 &   3.6 & Round\_Trip & 0.12  &    71 \\
 2015.79 & KEPLER-138   & 19082 &   2.5 &   1   & Round\_Trip & 0.07  &    91 \\
 2015.81 & GJ 3470      & 24896 &   5.8 &   4.9 & Round\_Trip & 0.24  &    60 \\
 2015.83 & HAT-P-18     & 36780 &   3.1 &  13.5 & Forward    & 0.022 &   123 \\
 2015.9  & WASP-76      & 33172 & 161.1 & 170.2 & Round\_Trip & 0.22  &    54 \\
 2015.93 & EPIC 2019125 & 17578 &   6.2 &   2.4 & Round\_Trip & 0.14  &    67 \\
 2015.94 & HAT-P-12     & 30330 &   9.2 &   3.8 & Forward    & 0.03  &   119 \\
 2016.05 & HAT-P-32     & 23257 &  11.1 &   5.8 & Forward    & 0.05  &   117 \\
 2016.1  & WASP-121     & 20865 &  12.5 &  11.7 & Forward    & 0.12  &    75 \\
 2016.11 & HAT-P-18     & 28312 &   6.6 &   5.5 & Forward    & 0.03  &   117 \\
 2016.15 & HAT-P-3      & 29747 &  16.2 &  16.5 & Forward    & 0.07  &    97 \\
 2016.17 & HAT-P-38     & 27871 &   5.2 &  19.5 & Forward    & 0.026 &   127 \\
 2016.19 & HD 149026    & 40531 &   4.8 &   3.4 & Round\_Trip & 0.7   &   103 \\
 2016.19 & HAT-P-26     & 29526 &   3.7 &   2.4 & Forward    & 0.06  &    92 \\
 2016.2  & EPIC 2019125 & 17515 &   3   &   5   & Round\_Trip & 0.14  &    63 \\
 2016.21 & HATS-7       & 20957 &   2.5 &   3.8 & Forward    & 0.02  &   143 \\
 2016.22 & HATS-7       & 20942 &   3.3 &   3.7 & Forward    & 0.02  &   132 \\
 2016.26 & HD 149026    & 40582 &   3.1 &   2.5 & Round\_Trip & 0.7   &    68 \\
 2016.28 & HD 97658     & 47835 &   2.2 &  33.9 & Round\_Trip & 1.4   &   132 \\
 2016.29 & WASP-29      & 33646 &   2.9 &   2.3 & Forward    & 0.11  &    69 \\
 2016.33 & HAT-P-26     & 29626 &   3   &   2.1 & Forward    & 0.06  &    97 \\
 2016.34 & 2MASS J23062 & 22034 &   3.6 &   3   & Round\_Trip & 0.027 &   145 \\
 2016.38 & EPIC 2019125 & 17614 &   5.1 &   4.5 & Round\_Trip & 0.14  &    66 \\
 2016.47 & WASP-80      & 19536 &   3.1 &   5.1 & Round\_Trip & 0.22  &    53 \\
 2016.49 & HAT-P-3      & 29628 &  17   &  17.5 & Forward    & 0.07  &    92 \\
 2016.51 & EPIC 2037710 & 16960 & 502.2 & 160   & Round\_Trip & 0.16  &   222 \\
 2016.62 & WASP-69      & 20934 &   3.9 &   6.9 & Round\_Trip & 0.63  &    47 \\
 2016.65 & HAT-P-38     & 27966 &   8.2 &  14.8 & Forward    & 0.026 &   133 \\
 2016.66 & WASP-52      & 30704 &   2.9 &   2.6 & Forward    & 0.035 &   159 \\
 2016.66 & WASP-39      & 28531 &   4.9 &   5   & Forward    & 0.035 &   108 \\
 2016.66 & HAT-P-12     & 30085 &   5.8 &   8.5 & Forward    & 0.03  &   121 \\
 2016.72 & WASP-63      & 26661 &   7   &   8.1 & Round\_Trip & 0.08  &    68 \\
 2016.75 & WASP-101     & 23949 &  21.7 &  12.3 & Forward    & 0.15  &    64 \\
 2016.76 & WASP-74      & 31595 &   4.9 &   6.2 & Forward    & 0.25  &    59 \\
 2016.77 & HAT-P-41     & 25061 &   5.3 &  11.9 & Forward    & 0.065 &    95 \\
 2016.79 & KELT-1       & 23460 &   4   &   2.5 & Round\_Trip & 0.097 &    65 \\
 2016.79 & HAT-P-41     & 24977 &   5.2 &   9.3 & Forward    & 0.065 &    89 \\
 2016.81 & WASP-67      & 29656 &   5.5 &   7.6 & Forward    & 0.037 &   113 \\
 2016.84 & WASP-76      & 33191 &  77.5 & 122.6 & Round\_Trip & 0.22  &    51 \\
 2016.86 & WASP-121     & 20929 &   4.9 &   4.8 & Forward    & 0.12  &    80 \\
 2016.87 & WASP-79      & 25289 &   2.1 &   3.3 & Forward    & 0.15  &    66 \\
 2016.88 & KELT-1       & 23455 &   6.9 &   7.3 & Round\_Trip & 0.097 &    73 \\
 2016.89 & KELT-1       & 23402 &   6.2 &   5.2 & Round\_Trip & 0.097 &    69 \\
 2016.89 & KELT-1       & 23444 &   5.8 &   4.7 & Round\_Trip & 0.097 &    70 \\
 2016.9  & K2-3         & 19933 &  78   &  34.2 & Round\_Trip & 0.18  &    60 \\
 2016.92 & K2-18        & 17539 &   7.6 &   1.6 & Round\_Trip & 0.14  &    63 \\
 2016.96 & HAT-P-32     & 23256 &   8.3 &   2.8 & Forward    & 0.05  &   144 \\
 2016.98 & HAT-P-7      & 31721 &   5.9 &   2.4 & Round\_Trip & 0.08  &    61 \\
 2016.99 & 2MASS J23062 & 22039 &   3.3 &  11.1 & Forward    & 0.027 &   167 \\
 2017.01 & HAT-P-7      & 31499 &   3   &   1.8 & Round\_Trip & 0.08  &    70 \\
 2017.01 & K2-18        & 17565 &   4.4 &   4   & Round\_Trip & 0.14  &    65 \\
 2017.02 & K2-3         & 19950 & 187.6 & 129.2 & Round\_Trip & 0.18  &    76 \\
 2017.03 & 2MASS J23062 & 22022 & 652.4 & 379.5 & Forward    & 0.027 &   384 \\
 2017.03 & HAT-P-18     & 28577 &   8.1 &   6.7 & Forward    & 0.03  &   120 \\
 2017.08 & HD 97658     & 47456 &   3.7 &  19.7 & Round\_Trip & 1.4   &   142 \\
 2017.1  & K2-18        & 17527 &   5.9 &   2.3 & Round\_Trip & 0.14  &    70 \\
 2017.1  & WASP-39      & 28363 &   3.7 &   3.4 & Forward    & 0.035 &   113 \\
 2017.14 & K2-3         & 20115 &   4.4 &   6.2 & Round\_Trip & 0.18  &    56 \\
 2017.17 & WASP-79      & 28110 &  42.7 &   2.7 & Forward    & 0.135 &    62 \\
 2017.27 & K2-3         & 20099 &   5   &   3.3 & Round\_Trip & 0.18  &    57 \\
 2017.28 & K2-18        & 17638 &   4.1 &   3.9 & Round\_Trip & 0.14  &    64 \\
 2017.29 & WASP-62      & 32905 & 158   &  56   & Forward    & 0.12  &    69 \\
 2017.3  & LHS 281      & 21461 &   5.7 &   3   & Round\_Trip & 0.2   &    93 \\
 2017.33 & WASP-74      & 31996 &   4.5 &   2.1 & Forward    & 0.25  &    66 \\
 2017.34 & WASP-6       & 15560 &   3.8 &   5.2 & Forward    & 0.06  &    96 \\
 2017.39 & USCO J161014 & 20067 & nan   & nan   & Forward    & 0.037 &   nan \\
 2017.43 & TYC 5530-179 & 28497 &   4.9 &   3.7 & Round\_Trip & 0.134 &    58 \\
 2017.44 & KEPLER-16    & 24216 &  19.8 &  33.7 & Round\_Trip & 0.097 &    61 \\
 2017.56 & WASP-17      & 33190 &   5.4 &  15.2 & Forward    & 0.033 &   113 \\
 2017.63 & KELT-7       & 17197 &   5.5 &   2.5 & Round\_Trip & 0.9   &    68 \\
 2017.73 & LHS 281      & 21470 &   5.7 &   5.4 & Round\_Trip & 0.2   &    57 \\
 2017.8  & KELT-7       & 17080 &   7.4 &   4.6 & Round\_Trip & 0.9   &    73 \\
 2017.87 & K2-3         & 20019 &   7.1 &   3.3 & Round\_Trip & 0.18  &    69 \\
 2017.88 & LHS 281      & 21480 &   6.4 &   3.4 & Round\_Trip & 0.2   &    62 \\
 2017.89 & LHS 281      & 21442 &   6.6 &   4.6 & Round\_Trip & 0.2   &    56 \\
 2017.9  & BD-02 5958   & 38142 &   5.4 &   3   & Round\_Trip & 0.365 &    64 \\
 2017.9  & LHS 281      & 21438 &   6.6 &   4.2 & Round\_Trip & 0.2   &    78 \\
 2017.91 & K2-18        & 17519 &   6.6 &   2.7 & Round\_Trip & 0.14  &    82 \\
 2017.94 & 2MASS J23062 & 28005 &   5.5 &   2.2 & Forward    & 0.02  &   151 \\
 2017.95 & GJ 3053      & 21073 &   7   &   3.2 & Round\_Trip & 0.14  &    61 \\
 2017.96 & BD-02 5958   & 38522 &   4.4 &   1.8 & Round\_Trip & 0.365 &    61 \\
 2018    & K2-3         & 20085 &  17.9 &   3.4 & Round\_Trip & 0.18  &    63 \\
 2018.04 & HIP 41378    & 49661 &   4.8 &   4.6 & Round\_Trip & 0.25  &    47 \\
 2018.12 & K2-3         & 19937 & nan   & 943.5 & Round\_Trip & 0.18  &  3847 \\
 2018.2  & WASP-121     & 32429 &   7.3 &   3.5 & Forward    & 0.073 &    70 \\
 2018.27 & WASP-127     & 26542 &   7.8 &   9.6 & Forward    & 0.17  &    71 \\
 2018.28 & KEPLER-79    &  6732 &   5.6 &  16.6 & Forward    & 0.015 &   282 \\
 2018.34 & HIP 41378    & 49718 &   7.9 &   6.6 & Round\_Trip & 0.25  &    40 \\
 2018.36 & K2-18        & 17544 & nan   & nan   & Round\_Trip & 0.14  &   nan \\
 2018.38 & HD 3167      & 46232 &   7.1 &   6.8 & Round\_Trip & 0.429 &    47 \\
 2018.45 & HD 106315    & 51569 & nan   & nan   & Round\_Trip & 0.213 &   nan \\
 2018.47 & HD 3167      &    18 & nan   & nan   & Round\_Trip & 0.429 &   nan \\
 2018.55 & 2MASS J23062 & 27416 &   7.3 &   8.3 & Forward    & 0.02  &   162 \\
 2018.55 & HD 3167      & 46606 &   7   &   8   & Round\_Trip & 0.429 &    53 \\
 2018.61 & GJ 1214      & 21621 &   8.2 &   7.3 & Round\_Trip & 0.12  &    63 \\
 2018.72 & GJ 9827      & 38835 & nan   & nan   & Round\_Trip & 0.365 &   nan \\
 2018.75 & BD-02 5958   & 38131 & 533.5 & 476.7 & Round\_Trip & 0.365 &   822 \\
 2018.85 & KEPLER-79    &  6678 &   5.6 &  11.3 & Forward    & 0.015 &   283 \\
 2018.9  & BD-02 5958   & 38276 &   3.9 &   3   & Round\_Trip & 0.365 &    59 \\
 2018.91 & HD 106315    & 52060 &   5.9 &   4.6 & Round\_Trip & 0.213 &    44 \\
 2018.97 & HD 106315    & 52148 &   5.1 &   4   & Forward    & 0.213 &    36 \\
 2018.99 & 2MASS J00041 & 23680 &   6.3 &   7.4 & Forward    & 0.022 &   134 \\
 2019.09 & HD 106315    & 52320 &   3.7 &   3.6 & Round\_Trip & 0.213 &    44 \\
 2019.09 & WASP-121     & 32395 &   5.8 &   4.1 & Forward    & 0.073 &    67 \\
 2019.45 & HD 3167      & 46254 &   6.5 &   4.3 & Round\_Trip & 0.429 &    55 \\
 2019.46 & TYC 5165-481 & 19739 &   4.9 &   7.4 & Round\_Trip & 0.22  &    57
\end{longtable}
\begin{tablenotes}
    \item [a] Measured maximum pixel fluence.
    \item [b] Median spectroscopic light curve precision over 16 channels.
\end{tablenotes}
\end{ThreePartTable}
\end{center}

\end{document}